\documentclass{ws-ijmpa}

\begin{document}

\markboth{B. Krippa, M. C. Birse, J.A.McGovern and N. R. Walet}
{Exact renormalization group and many fermion systems}

\catchline{}{}{}{}{}
%
%%%%%%%%%%%%%%%%%%%%%%%%%%%%%%%%%%%%%%%%%%%%%%%%%%%%%%%%%%%%%%%%%%%%

\title{EXACT RENORMALIZATION GROUP AND MANY-FERMION SYSTEMS\\
}

\author{B. Krippa$^{1,2}$, M. C. Birse$^{1}$, J. A. McGovern$^{1}$, N. R. Walet$^{2}$}
\address{$^{1}$Department of Physics and Astronomy, University of Manchester,
\\
Manchester, M13 9PL, UK\\
 $^{2}$Department of Physics, UMIST
P.O.Box, Manchester, M60 1QD, UK 
}

\maketitle

%\pub{Received (Day Month Year)}{Revised (Day Month Year)}

\begin{abstract}

The exact renormalization group method is applied
to many-fermion systems with short-range attractive forces.
The strength of the attractive fermion-fermion interaction is
determined from the vacuum scattering length.
A set of approximate flow equations is derived including 
fermionic bosonic fluctations. The numerical solutions show 
a phase transition to a gapped phase.  The inclusion
of bosonic fluctuations is found to be significant only in
the small-gap regime.

\keywords{Superfluidity; Pairing, Renormalization Group; Phase Transition.}
\end{abstract}
\vspace{10pt}

Attractive forces between fermions play a crucial role in many areas of
many-body physics. In a system of fermions, any attraction 
 can cause such particles to form correlated 
``Cooper pairs''  so that a phase transition occurs.
 The order parameter for this is the
energy gap, which appears in the fermion spectrum. Within such
systems we can identify a weak-coupling regime, with no
two-body bound states, which corresponds to
Bardeen-Cooper-Schrieffer (BCS) superconductivity and a
strong-coupling regime, with a deeply bound two-body state,
which manifests itself in a form of Bose-Einstein Condensation (BEC). 

A new and promising way to treat many-fermion systems with these
different regimes is provided by the Exact Renormalization Group
(ERG)\cite{Wi74}. This has been successfully applied to a variety of other  
systems in particle and condensed-matter physics\cite{BTW02}.
The goal of the ERG approach is to construct the Legendre transform of
the effective action, $\Gamma$, which generates the 1PI Green's functions. 
One introduces an artificial renormalisation group flow for 
$\Gamma$, depending on a momentum scale $k$, and defines a
running effective action by integrating over components of the 
fields with $q < k$.  The RG trajectory then interpolates between the 
classical action of the underlying theory (at large $k$), and the 
full effective action (at $k=0$)\cite{BTW02}.

Here we study a system of fermions interacting through an attractive 
two-body contact potential.  We take as our starting point an EFT that 
describes the $s$-wave interaction of two fermions with 
scattering length $a_0$.
Since we are interested in the appearance of a gap in the fermion
spectrum, we introduce a boson field whose vacuum 
expectation value (VEV) describes that gap and so acts as the 
corresponding order parameter. We take the following
Ansatz for $\Gamma$:
\begin{eqnarray}
&&\Gamma[\psi,\psi^\dagger,\phi,\phi^\dagger,\mu,k]=\int d^4x\,
\left[\phi^\dagger(x)\left(Z_\phi\, i \partial_t 
+\frac{Z_m}{2m}\,\nabla^2\right)\phi(x)-U(\phi,\phi^\dagger)\nonumber\right.\\
&&\qquad\left.+\psi^\dagger\left( Z_\psi (i \partial_t+\mu)
+\frac{Z_M}{2M}\,\nabla^2\right)\psi
- Z_g\left(\frac{i}{2}\,\psi^{\rm T}\sigma_2\psi\phi^\dagger
-\frac{i}{2}\,\psi^\dagger\sigma_2\psi^{\dagger{\rm T}}\phi\right)\right].
\label{eq:Gansatz}
\end{eqnarray}
Here $M$ is the mass of the fermions in vacuum and $m$ is 
taken to be $2M$. 

We expand the potential $U$ about its minimum, $\phi^\dagger=\phi=\Delta$ 
where $\Delta$ is the fermion energy gap. To quartic order this has the form
\begin{equation}
U(\phi,\phi^\dagger)= u_0+ u_1(\phi^\dagger\phi-\Delta^2)
+{\textstyle\frac{1}{2}}\, u_2(\phi^\dagger\phi-\Delta^2)^2.
\end{equation}
In the RG evolution, we start at high $k$ from an almost free bosonic action
where the VEV of $\phi$ is zero. When $k$ is lowered, 
we expect that at some scale the system undergoes 
a transition to a phase with spontaneously broken $U(1)$ symmetry 
and a nonzero fermion gap $\Delta$.  

The evolution equation for $\Gamma$ in the ERG has a straightforward
one-loop structure\cite{BTW02} and in the case of constant $\mu$ it
has the form
\begin{equation}
\partial_k\Gamma=-\frac{i}{2}\ {\rm Tr} \left[
(\Gamma ^{(2)}_{BB}-R_B)^{-1}\,\partial_{k} R_B\right]
+\frac{i}{2}\ {\rm Tr} \left[
(\Gamma ^{(2)}_{FF}-R_F)^{-1}\,\partial_{k} R_F\right].
\label{eq:Gamevol}
\end{equation}
Here $\Gamma ^{(2)}_{FF(BB)}$ is the matrix of the second
derivatives of the effective action with respect to the
fermion (boson) fields and $R_{B(F)}$ is a matrix containing the
corresponding boson (fermion) regulators.
The usual mean-field results can be recovered if we 
include fermion loops only (the second term of this equation).
By evaluating the loop integrals in (\ref{eq:Gamevol})  
we get a set of coupled equations for the coefficents $u_n$
and $Z_\phi$ at constant $\mu$. Details can be found 
in our paper\cite{BK1}. The initial conditions 
are obtained by assuming that in vacuum our theory reproduces the scattering 
length $a_0$, and that the necessary subtractions are identical in matter and
in vacuum. 

\begin{figure}
\begin{center}
\includegraphics[width=8.5cm,  keepaspectratio,clip]{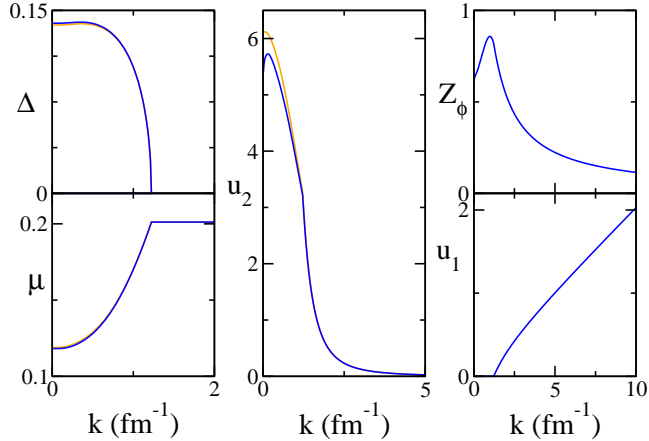}
\end{center}
\caption{\label{fig:running}Numerical solutions to the evolution equations
for the  relevant parameters for infinite $a_0$ and $p_F=1.37$~fm, starting 
from $K=16$~fm$^{-1}$.
The black line shows the full evolution with
both fermion and boson loops; the grey line the result with fermion loops only.
All quantities are expressed in appropriate powers of fm$^{-1}$.}
\end{figure}

Although our results can be applied to many systems, 
for definiteness we concentrate on parameter values relevant  
to neutron matter: $M=4.76$~fm$^{-1}$, $p_F=1.37$~fm$^{-1}$, 
and large two-body scattering lengths ($|a_0| > 1$~fm). 
We have checked the dependence of our results on the starting 
scale $K$ and find that this is undetectable as long as $K>5$~fm$^{-1}$.
Some typical solutions for the evolution equations are given in 
Fig.~\ref{fig:running}, for the case of infinite $a_0$. We compare  
two different approximation schemes, one where we keep fermion loops only, 
and one where we include boson loops as well and we allow $Z_\phi$ to run.
For large values of $k$ the system remains in the symmetric phase. 
At $k\simeq 1.2$~fm$^{-1}$, the first derivative
of $U$ at $\phi=0$ vanishes and the system enters the broken phase. 

We find that the contributions of boson loops are small. 
Indeed in the symmetric phase they have essentially no effect.
Below the transition they do become visible, particularly in $u_2$.
However their effects on the gap are even smaller, at most $\sim 1\%$ 
for $1/|p_Fa_0|<1$. In this region, the main effect of the bosons is a small
enhancement of the gap related to the reduction in $u_2$.

We have also examined the behaviour of the gap over a wider range
of $1/(p_{F}a_{0})$. The overall picture is the same as in Marani et 
al.\cite{Pis} for fermion loops only, with a crossover from BCS pairing (with  
$\mu\simeq \epsilon_F$) for $1/(p_{F}a_{0})<0$ to BEC (with $\mu < 0$)
for $1/(p_{F}a_{0})>0$. Deviations from mean-field behaviour are present in 
the BEC region and become increasingly noticable for weaker couplings or lower 
densities.

In the case of neutron matter, we find gaps of the order of 30~MeV.
There is a simple explanation for such unrealistically large values\cite{Fayans}. 
For weak couplings, the gap is proportional to 
$\exp\left(-(\pi/2)\cot\delta(p_F)\right)$.
For nucleon-nucleon scattering, $\cot\delta$ increases quickly
with momentum, resulting in a reduction of the gap.
We therefore expect that inclusion of the 
effective range should capture this physics.

There a number of improvements which should be made to our approach. 
Adding an effective range is clearly an important one. Notes that
this will also require inclusion of the three-body forces to respect the 
reparametrization invariance\cite{BK2}. The running of the boson kinetic 
mass ($Z_m$), the fermion renormalization constants and the ``Yukawa'' 
coupling are needed for a full treatment of the action (\ref{eq:Gansatz}).
Beyond that we would also like to treat explicitly particle-hole channels
(RPA phonons).

\end{document}